\documentclass[prl,amsmath,amssymb,twocolumn,showpacs,floatfix,superscriptaddress]{revtex4}
\usepackage{amsmath,amsfonts}
\usepackage{graphicx}
\usepackage{color}
\usepackage{enumerate}

\newcommand{\e}{\mathrm{e}}

\newcommand{\bI}{\mathbf{I}}
\newcommand{\bS}{\mathbf{S}}

\newcommand{\bB}{\mathbf{B}}
\newcommand{\mean}[1]{\langle #1 \rangle}

\newcommand{\up}{\uparrow}
\newcommand{\dw}{\downarrow}

\DeclareMathOperator{\arctanh}{arctanh}

\newcommand{\ps}{}

\begin{document}


\title{Interplay between classical magnetic moments and superconductivity in quantum one-dimensional conductors: toward a self-sustained topological Majorana phase}

\author{Bernd Braunecker}
\affiliation{Departamento de F\'{\i}sica Te\'{o}rica de la Materia Condensada,
Centro de Investigaci\'{o}n de F\'{\i}sica de la Materia Condensada,
and Instituto Nicol\'{a}s Cabrera,
Universidad Aut\'{o}noma de Madrid, E-28049 Madrid, Spain}
\author{Pascal Simon}
\affiliation{Laboratoire de Physique des Solides, CNRS UMR-8502,
             Universit\'{e} de Paris Sud, 91405 Orsay Cedex, France}

\date{\today}



\begin{abstract}
We study a one-dimensional (1D) interacting electronic liquid coupled to a 1D array of classical magnetic moments and to a superconductor. We show that at low energy and temperature the magnetic moments and the electrons become strongly entangled and that a magnetic spiral structure emerges. For strong enough coupling between the electrons and magnetic moments,
the 1D electronic liquid is driven into a topological superconducting phase supporting Majorana fermions without any fine-tuning of external parameters. Our analysis applies at low enough temperature to a quantum wire in proximity of a superconductor when the hyperfine interaction between electrons and nuclear spins is taken into account, or to a chain of magnetic adatoms adsorbed on a superconducting surface.
\end{abstract}


\maketitle


{\em Introduction.}
The interaction between localized magnetic moments and delocalized electrons
contains the essential physics of many modern condensed matter systems.
It is on the basis of nuclear magnets \cite{froehlich:1940}, heavy fermion materials
of the Kondo-lattice type \cite{tsunetsugu:1997} or
ferromagnetic semiconductors \cite{ohno:1992,ohno:1998,dietl:1997,koenig:2000}.
It often leads to new intricate physics and rich phases diagrams already when the magnetic moments behave classically.
Electron systems interacting with nuclear spins through the hyperfine interaction or magnetic adatoms with large magnetic moments arranged in some array on a metallic surface enter into this class.

In 1D, the interactions between  the nuclear spins and electrons lead to  dramatic effects: below a cross-over temperature $T^*$, a new exotic phase of matter in which the nuclear magnets are strongly tied to the electrons naturally emerges \cite{braunecker:2009a,braunecker:2009b}.
In this phase, the nuclear spins form a helical magnetic structure caused by the effective Ruderman-Kittel-Kasuya-Yosida (RKKY) interaction \cite{RKKY} mediated by the electron system. The feedback of this nuclear Overhauser field on the electron system entirely restructures the electronic states in that it opens a gap
in one half of the elementary low-energy modes.
The remaining electronic degrees of freedom remain gapless and form a
quasi helical Luttinger liquid  with strong analogies \cite{braunecker:2012} with the edge states of the 2D quantum spin Hall
effect \cite{kane-mele,wurzburg}.
Due to the mutual feedback this order of strongly coupled electrons and nuclear spins is stable below a temperature $T^*$,
and electron-electron interactions substantially enhance the stability \cite{braunecker:2009b}.
Recent transport measurements in cleaved edge overgrowth GaAs quantum wires found a reduction of the conductance by a factor of two below $T<100~{\rm mK}$ independently of the density or applied magnetic field, consistent with this theory \cite{zumbuhl}.

We stress that the mechanism behind this emergent helical structure is general, the essential ingredient being the RKKY interaction.
Therefore, the same mechanism can apply if the nuclear spins are replaced by classical magnetic moments forming a 1D lattice (not necessarily a regular one) such as magnetic adatoms on top of a metallic surface \cite{menzel}.

\begin{figure}
\begin{center}
	\includegraphics[width=\columnwidth]{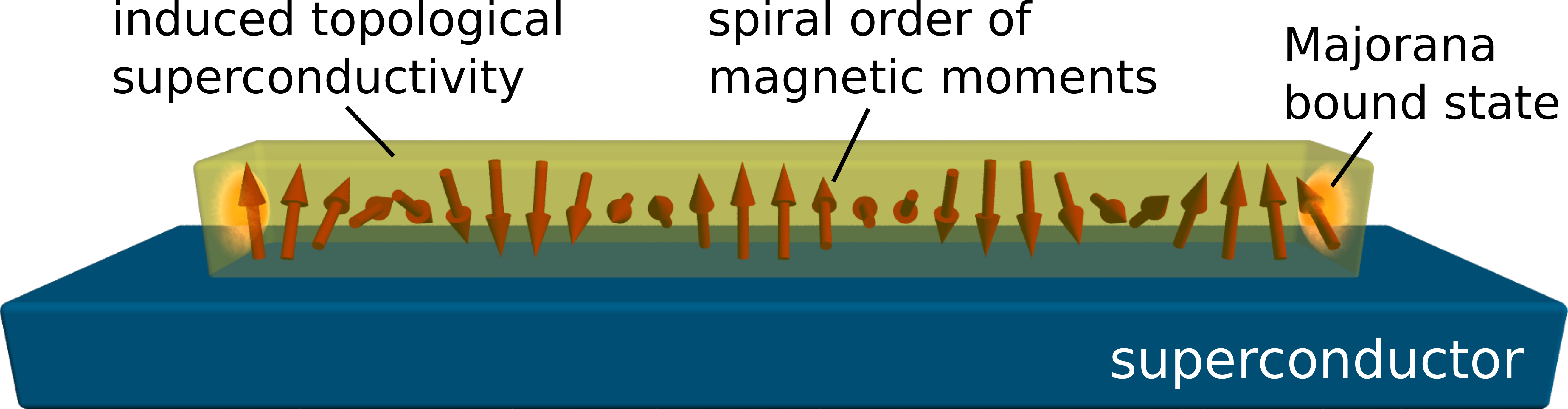}
	\caption{\label{fig:system}
(Color online)	A conductor with large magnetic moments placed on top of a superconductor.
	Topological superconductivity, Majorana bound states, and a spiral order
	of the magnetic moments emerge from a self-organization of the coupled systems
	of electrons and magnetic moments.
	}
\end{center}
\end{figure}

When a finite-sized helical liquid is put in proximity of a s-wave superconductor, Majorana states can emerge at both ends \cite{kane_RMP} (see Fig. \ref{fig:system}). This is the case for a quantum wire in presence of spin-orbit coupling and a Zeeman term \cite{lutchyn,oreg} where some possible signatures of Majorana fermion physics have been recently reported experimentally  \cite{mourik,deng,heiblum}.
The helical liquid, up to a gauge transformation \cite{braunecker:2010}, can also be obtained by coupling electrons to a spiral magnetic field (like the intrinsic nuclear Overhauser field \cite{braunecker:2009b}),
or by manufacturing an external rotating magnetic field \cite{flensberg:2012}.
They can appear in rare-earth compounds exhibiting coexisting helical magnetism and
superconductivity \cite{morpurgo}, or emerge by arranging magnetic adatoms in 1D arrays on the surface of a superconductor \cite{akhmerov,yazdani}.

Since induced or intrinsic superconductivity entirely restructures the electron system, it is not a priori obvious whether the helical entangled states remain stable. In this work we provide such an investigation, carefully taking also into account electron-electron interactions and disorder.
\ps{The underlying physics is the following:
The RKKY interaction between the local moments has a
strong $2k_F$ component (with $k_F$ the Fermi wave vector), such that the local moments tend to be
opposite at length $\lambda_F/4$, with $\lambda_F=2\pi/k_F$. The RKKY energy can thus be minimized
if the local moments form a spiral \cite{braunecker:2009a,braunecker:2009b}. 
This spiral acts back on the electron as the effective spiral Zeeman field
required, together with the induced superconductivity, to drive the system into the topological superconductivity phase. 
Yet the opening of the superconducting gap as well as renormalizations by interactions and disorder modify 
again the RKKY interaction. Taking this self-consistently into account, however, we demonstrate that a self-stabilizing 
topological phase supporting Majorana bound states naturally emerges.
We emphasize that this topological phase
requires no fine-tuning. It is an intrinsic effect and constitutes the thermodynamic ground state of the system.}


{\em Model Hamiltonian.}
We consider a 1D conduction electron liquid in proximity of a s-wave superconductor. The 1D electrons are further coupled to an array of magnetic moments. Such a generic system is of the 1D Kondo-lattice type and described by the Hamiltonian
\begin{equation} \label{eq:H}
	H =
	H_{el}
	+ \sum_{x_i} A_0 \bS_{x_i} \cdot \bI_{x_i}.
\end{equation}
The first term, $H_{el}$ describes the electron system in the induced or intrinsic
superconducting state and also includes electron-electron interactions. The second term describes the coupling between the electron spins
$\bS_{x_i}$ and the magnetic moments $\bI_{x_i}$, weighted by a coupling constant $A_0$.
The positions of the magnetic moments $x_i$ lie on a 1D chain. We assume that the distances
between neighboring moments fulfill $|x_{i+1} - x_i| \ll \lambda_F$.
Yet it is not required that the $x_i$ lie
on a regular 1D lattice. The operators $\bS_{x_i}$ are a tight binding representation of
the electron spin operator in a region of dimension $\delta x \ll \lambda_F$ centered about $x_i$.
The quantities $\bI_{x_i}$ are either very large spins $I$ or composites of a large number $N_\perp$
of individual magnetic moments $I$  locked to a parallel configuration in some small
volume at $x_i$ (such as nuclear spins in the transverse section of a quantum wire \cite{braunecker:2009b}). The former case is treated with $N_\perp=1$.
We assume that $\bI_{x_i}$ is normalized with respect to $N_\perp$ such that $A_0$ represents the
interaction constant between an electron and one of the $N_\perp$ individual moments $I$ of $\bI_{x_i}$.
\ps{It has been shown numerically in \cite{yazdani} and further justified in \cite{pientka} that this simple model captures qualitatively the behavior of a regular array of magnetic adatoms
adsorbed on a superconducting surface (see also \cite{nagaosa} for the 2D case).}

The magnetic coupling $A_0$ also provides an effective RKKY \cite{RKKY} interaction
and dynamics for the magnetic moments. This interaction is carried over the
response of the electron system to magnetic perturbations and consists in an
almost instantaneous long-ranged Heisenberg interaction between the magnetic moments
$J(x_i-x_j) = A_0^2 a_I^2 \chi(x_i-x_j)/2$, where $\chi(x)$ is the electron spin susceptibility and $a_I$ the lattice spacing between
the moments $I$.
Note that generally $J$ can be anisotropic.

To analyze the Hamiltonian \eqref{eq:H}, we resort to a Born-Oppenheimer decoupling as in \cite{braunecker:2009b}
which relies on the fact that
the magnetic moments have a much slower dynamics than the electrons.
Since the large moments $\bI_{x_i}$ allow a (quasi)classical
description, the terms $A_0 \bI_{x_i}$ act like a local quasi-static Zeeman field
$\bB_\text{eff}(x_i) = A_0 \mean{\bI_{x_i}}$ on the electron spins
$\bS_{x_i}$.
The resulting effective theory is, therefore, expressed by the pair of Hamiltonians
 \cite{simon:2007,braunecker:2009a,braunecker:2009b}
\begin{align}
\label{eq:H_el_eff}
	H_{el}^{\text{eff}}
	&= H_{el} + \sum_{x_i} \bB_\text{eff}(x_i) \cdot \bS_{x_i},
\\
\label{eq:H_m_eff}
	H_{m}^{\text{eff}}
	&= \sum_{x_i,x_j} J(x_i-x_j) \bI_{x_i} \cdot \bI_{x_j}.
\end{align}
While $H_{el}^{\text{eff}}$ acts only
on electrons and $H_m^{\text{eff}}$ only on the magnetic moments, we stress that
both Hamiltonians are strongly coupled since $\bB_\text{eff}$ depends on the state
of the magnetic moment system, and $J$ on the electron state. A characterization of the
physics described by Eqs. \eqref{eq:H_el_eff} and \eqref{eq:H_m_eff} must rely on a
fully \emph{self-consistent} approach, similar to the treatment of Refs.
\cite{simon:2007,braunecker:2009a,braunecker:2009b}.

{\em Susceptibility in the non-interacting case.}
Without superconductivity, the magnetic moments order in a spiral at low temperature due to
the self-organization
of the coupled systems \cite{braunecker:2009a,braunecker:2009b}, leading to an effective spiral field of amplitude
$B_\text{eff} = A_0 I$ and
spatial period $\lambda_F/2=\pi/k_F$.
Since the spiral field rotates in
a plane defining, for instance, the spin $(x,y)$ plane, the RKKY interaction is governed by the
transverse spin susceptibility, defined as \cite{braunecker:2009b}
$\chi^\perp(q) = -i \int_0^\infty dt \int dx \ \e^{i q x} \ \mean{[S_+(x,t),S_-(0,0)]}$,
for momenta $q$ with $S_+ = \psi_\up^\dagger \psi_\dw$, $S_- = S_+^\dagger$, and $\psi_{\sigma=\up/\dw}(x)$
destroying an electron of spin $\sigma$ at position $x$ ($\hbar=1$ throughout the paper).

We first neglect electron interactions.
Without superconductivity, $H_{el}$ is then equivalent, by a gauge transformation \cite{braunecker:2010}, to a one-channel conducting wire with spin-orbit interaction in a magnetic field. The resulting band structure consists
of two spin-mixing bands $\alpha=\pm$ with dispersions $E_{k}^\alpha = \epsilon_{k} + \alpha \sqrt{B_\text{eff}^2 + v_F^2 k^2}$, where $\epsilon_k$ is the single-particle dispersion and $v_F$ the Fermi velocity. At $k=0$ both bands are separated by a gap of amplitude $2 B_\text{eff}$, such that
when the Fermi level lies within this gap, only band $\alpha=-$ remains conducting and forms a quasi-helical
conductor with roughly opposite spins at opposite Fermi points $k_{F-}$ (when undoing the gauge transformation
$k_{F-} \approx k_F$).
In the presence of superconductivity, the proximity induced gap $\Delta_S$ is expanded in the $\alpha=\pm$ eigenbasis and
results in two triplet pairing terms within the `$+$' and `$-$' bands,
$\Delta = \Delta_{++}=\Delta_{--}$, and a singlet term $\Delta_{+-}$
that mixes both bands. A similar expansion holds for the transverse electronic susceptibility, which we
write as
$\chi^{\perp}(q)=\chi^{\perp}_{--}(q)+\chi^{\perp}_{++}q)+\chi^{\perp}_{+-}(q)$.
Since both bands are separated by a gap, $\chi^\perp(q)$ is
dominated  by the gapless `$-$' band such that $ \chi^{\perp}(q)\approx \chi^{\perp}_{--}(q)$.
The analytical form of $\chi^\perp_{--}(q)$ is derived in the supplementary material.
It has a deep minimum of the normal state susceptibility at $q=2k_F$ (see Fig. 2).
For $\Delta_{--} \ll v_F k_F$, we find (see supplement) that it is well described by
\begin{equation}\label{eq:chi_perp}
	\chi^\perp_{--}(q)
	\approx
	\frac{-1}{2\pi v_F}
	\Biggl[
		\ln\left(\frac{2 v_F k_F}{\Delta}\right)
		-
		\frac{1}{\kappa}
		\arctanh(\kappa)
	\Biggr],
\end{equation}
with $\kappa = v_F (q-2k_F) / \sqrt{4 \Delta^2 + v_F^2 (q-2k_F)^2}$.
Since $\chi^\perp_{--}(q)$ maintains the deep minimum at $2k_F$ characteristic of the normal state, this confirms
on a mean field level
the consistency of the assumption of the spiral effective field, while stability under
fluctuations will be considered below. We also note that the minimum can only be
approximated by a Lorentzian of half-width $\Delta$ for $v_F|k-2k_F|\ll \Delta$.
\begin{figure}
\includegraphics[width=\columnwidth]{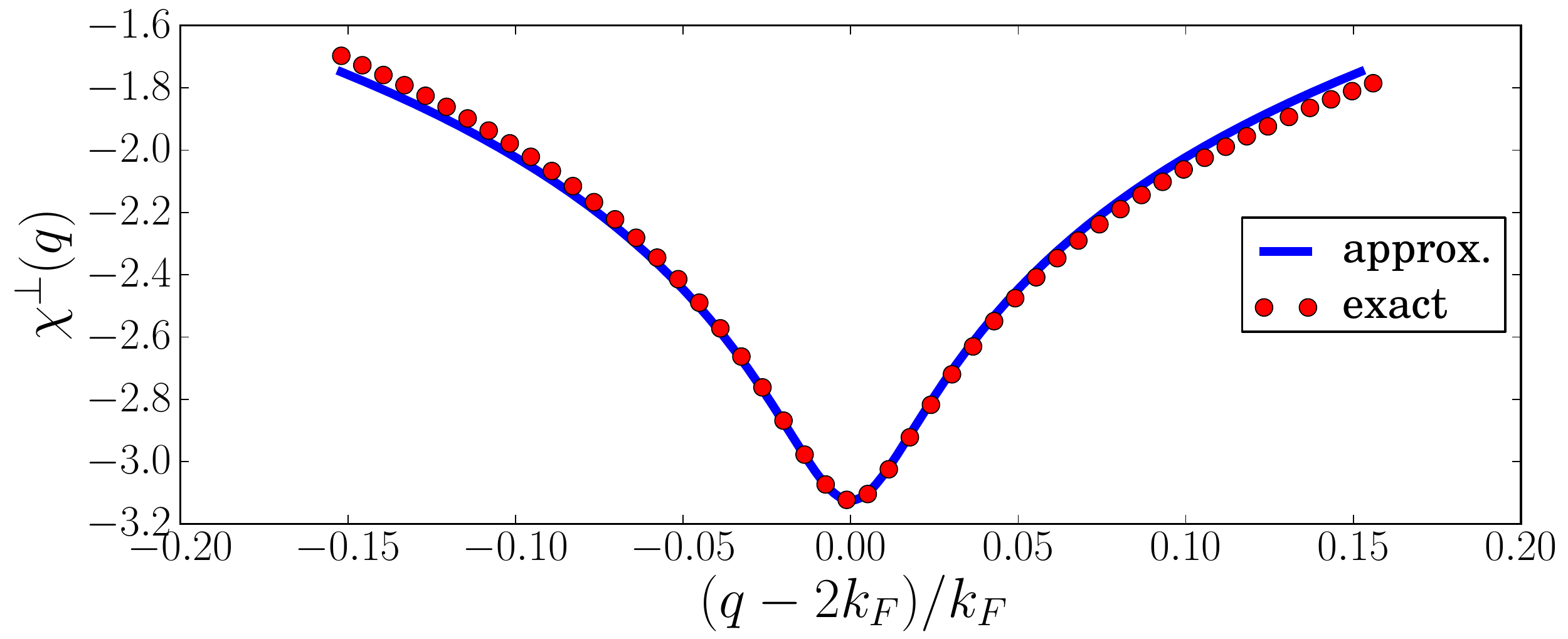}
\caption{\label{fig:chi_noninteracting}
	(Color online) Transverse susceptibility $\chi_{--}(q)$ for the non-interacting quasi-helical conductor with induced superconductivity.
 }
\end{figure}

{\em Susceptibility in the interacting case.}
In 1D conductors, electron-electron interactions often are detrimental.
To include them, we first linearize the spectrum in the absence of the both the proximity induced gap and of the spiral magnetic field. Our treatment is therefore valid only at low energy. We then use the standard approach of bosonizing the low-energy electronic Hamiltonian \cite{giamarchi} and incorporate
the pairing term and the spiral magnetic field in the bosonized Hamiltonian. The resulting low-energy model  becomes $H_{el}=H^0_{c}+H^0_{s}+H_{B}+H_{P}$ with
\begin{equation} \label{eq:H_0}
	H^{0}_{\nu=c,s}
	=
	\int \frac{dx}{2\pi} \left[
		\frac{v_\nu}{K_\nu} (\nabla\phi_\nu(r))^2 + v_\nu K_\nu (\nabla\theta_\nu(r))^2
	\right],
\end{equation}
\begin{equation} \label{eq:H_B}
	H_{B}=
	\int \frac{dx}{2\pi a} \, B_\text{eff}
	\cos\bigl(\sqrt{2}(\phi_c+\theta_s)\bigr),
\end{equation}
and
\begin{equation} \label{eq:H_P}
	H_{P}=
	\int dx \frac{2}{\pi a} \, \Delta
	\left[
		\sin\bigl(\sqrt{2}(\theta_c+\phi_s)\bigr)+\sin\bigl(\sqrt{2}(\theta_c-\phi_s)\bigr)
	\right].
\end{equation}
In Eq. \eqref{eq:H_0}, $K_{c,s}$ are the Luttinger Liquid  parameters for the charge and spin density fluctuations,
$v_{c,s}$ is the charge and spin density wave velocities, and $a\sim a_I$ is a short distance cutoff.
The non-interacting case
is described by $K_c = K_s = 1$ and $v_c = v_s = v_F$. Repulsive electron-electron interactions
lead to $0 < K_c < 1$.
If the spin SU(2) symmetry is preserved $K_s = 1$ otherwise $K_s > 1$. Strictly speaking, this symmetry is broken here by $B_\text{eff}$, yet only weakly such that $K_s\approx 1$ \cite{suhas:2007,egger}. The $\cos\sqrt{2}(\phi_c+\theta_s)$ term has a scaling dimension $(K_c+K_s^{-1})/2$ and the $\cos\sqrt{2}(\theta_c\pm\phi_s)$ terms the scaling dimension $(K_c^{-1}+K_s)/2$. In the non-interacting case both scaling dimensions are equal to 1
indicating the absence of any renormalization. For $1/3<K_c<1$ and $K_s=1$, both the pairing term and effective magnetic field terms are relevant, however, $B_\text{eff}$ dominates.

To make progress, we assume that $B_\text{eff}\gg \Delta$ (for strong repulsive interactions, one can relax this condition). 
\ps{This ensures that conduction modes unaffected by the opening of the $B_\text{eff}$ gap are helical, and that
the induced superconductivity in this helical conductor is topological \cite{lutchyn,oreg,suhas}.}
We proceed in a 2-step renormalization group (RG) analysis.
The coupling $B_\text{eff}$ reaches the strong coupling first, which opens a gap in the spectrum by pinning the field $\phi_c+\theta_s$ to a constant \cite{braunecker:2009b}.
This physics is best accessed by introducing the fields $(\phi_+,\theta_+,\phi_-,\theta_-)$ related to the original fields by a unitary transform (see the supplemental material), with the $\pm$ fields corresponding to the previous $\pm$ bands of the non-interacting case. In this new basis, the effective Zeeman term gives a simple dominant sine-Gordon term $B_{\rm eff}\cos(2\phi_+)$. Hence the gap opens only in the `$+$'
sector, and from the strong coupling limit we can estimate it to be
$B^*_{\rm eff}\sim (v_F/a) (aB_\text{eff}/v_F)^{2/(4-K_c-K_s^{-1})}$,
coinciding with $B_\text{eff}$ in the non-interacting limit and increasing as a power law otherwise.
After projecting out the gapped sector, the remaining Hamiltonian
takes the simpler form
\begin{equation}\label{eq:H_0-}
	H_-=H_0^- + \int dx \frac{2}{\pi a_B} \, \Delta' \sin(2\theta_-),
\end{equation}
where $H_0^-$ is given by Eq. \eqref{eq:H_0} with $K_\nu=K_-=2K_c/(1+K_cK_s)$, $v_-=v_cK_s^{-1}+v_sK_c$. Note that the effective bandwidth in this resulting `$-$' sector is now determined by $B_{\rm eff}$ such that $a_B \sim v_+/B_{\rm eff}$ replaces the UV cutoff $a$.
The pairing term  $\Delta$ is now replaced by the renormalized value
$\Delta'\lesssim \Delta$. The Hamiltonian \eqref{eq:H_0-} is nothing but the bosonized version of a spinless electronic chain in proximity of a superconductor.
which can can be  studied by our second step RG procedure.
As shown in Ref. \cite{suhas}, strong electron interactions further
renormalize the value of $\Delta'$, usually to $\Delta^{*} \ll \Delta'$.
However, by noticing that  $K_-$ also grows under the RG, we can refermionize
the system at a length scale $\tilde l$  defined by $K_-(\tilde l)=1$ \cite{suhas}. The refermionized
Hamiltonian is nothing but a {\it non-interacting} 1D triplet superconductor with a renormalized pairing
gap $\Delta^{*}$. The susceptibility is, therefore, given by Eq. \eqref{eq:chi_perp} with the replacements
 $\Delta$ by $\Delta^{*} \ll \Delta$
and $v_F$ by $v_-$. Repulsive electron interactions  therefore increase $B_{\rm eff}^*$ while decreasing $\Delta^*$.


{\em Stability analysis.}
In the previous analysis, we have found that
the RKKY interaction remains strongly peaked at $q=2k_F$ in the presence of the superconductivity, such that the
magnetic moments tend to form the spiral order.
Through the opening of the gap in the electron system by this self-organization, this $2k_F$ ($\lambda_F/2$)
spiral is strongly favored energetically.
To investigate its
stability, we perform a magnon analysis of $H_m^{\rm eff}$ similar to Ref. \cite{braunecker:2009b} analyzing the low-energy
fluctuations of the moments $\bI_{x_i}$.
If the $\bI_{x_i}$ are regularly spaced and composed by $N_\perp$ individual moments of size $I$ that are ferromagnetically locked to each other, the magnetization $m$ per site (normalized to $0 \le m \le 1$) reads \cite{braunecker:2009b}
$m = 1 - \frac{1}{IN N_\perp} \sum_q \frac{1}{\e^{\omega_q/k_B T}-1}$,
with $T$ the temperature, $k_B$ the Boltzmann constant, $N$ the number of sites of the chain, and the magnon dispersion
$\omega_q = 2I(J_{q+2k_F}-J_{2k_F})/N_\perp$ with $J_q = A_0^2 a_I \chi_{--}(q)$.
The order is stable as long as
$m \approx 1$, and we denote the crossover temperature at which the order disappears by $T^*$.
For an  infinite chain, $T^*=0$
and order cannot be stable. For finite systems of length $L = N a_I$,
this lowest mode is
cut off, leading to $T^*>0$.
Contrarily to the normal state case \cite{braunecker:2009b}, the RKKY interaction $J_q$ is determined
by $\Delta^{*}$ but not by $T$ (provided that $k_B T \ll \Delta^{*}$).
Hence $J_q$ remains invariant under temperature changes,
and $T^*$ becomes a function of the remaining scales characterizing $\omega_q$, notably the
cutoff scale
$\omega_{2\pi/L} \approx 2I (2\pi/L)^2 dJ_{q+2k_F}/d(q^2)|_{q=0}$, and the mean-field scale
$\omega_\infty = 2I |J_{2k_F}|/N_\perp$.

At very short system sizes such that only very few $q$ values
fall into the dip of the RKKY interaction, we have $\omega_q \approx \omega_\infty$ for almost all
terms in the $q$ summation, and we obtain $m \approx 1 - [\e^{\omega_\infty/k_B T}-1]^{-1}/I N_\perp$,
which leads to a vanishing $m$ at temperatures exceeding
\begin{equation} \label{eq:T*}
	k_B T^* = 2 I^2 |J_{2k_F}|
	=  A_0^2 I^2 a_I [\ln(2 v_- k_F/\Delta^{*})-1] / \pi v_-.
\end{equation}
Quite remarkably, this result remains accurate up to large system lengths in which the
approximation $\omega_q \approx \omega_\infty$ is no longer valid for most $q$.
Indeed, let us consider the $L \to \infty$ limit, in which the sum to calculate $m$
is dominated by the $q=2\pi/L$ term, $m \approx 1 - [\e^{\omega_{2\pi/L}T}-1]^{-1}/I N N_\perp$.
The associated temperature scale is $k_B T_{2\pi/L} = I N \omega_{2\pi/L} \sim 1/L$.
The length effects become influential only when $T^* \sim T_{2\pi/L}$, i.e., at the length
$L^*\approx \pi^2 v_-^2/\{3 a_I (\Delta^*)^2 [\ln(2v_- k_F/\Delta^*)-1]\} \sim (\xi^*)^2/a_I$.
For systems with $L > L^*$, the crossover temperature lies
between $T^*$ and $T_{2\pi/L}$, and decays with $L$ not faster than $T_{2\pi/L} \sim 1/L$.
However, since usually $\xi^*/a_I \gg 1$, $T^*$ remains $L$ independent far into the regime $L \gg \xi^*$
required for obtaining isolated Majorana bound states.
We notice that the unrenormalized $IA_0$ in Eq. \eqref{eq:T*} must not to be confused with the renormalized $B^*_\text{eff} \gg IA_0$ seen by the electrons. The topological phase requires $B^*_\text{eff} > \Delta^*$, yet within this situation both
$IA_0 \gg \Delta^*$ and $IA_0 \ll \Delta^*$ are possible.

{\em Disorder.} We have assumed so far that the system is free of disorder.
Since the RKKY magnetic interaction between the magnetic moments
is mainly dominated by $J_{2k_F}$, this remains the case even if the magnetic adatoms do not form a regular array. More problematic is the  disorder in the electronic part.  Semiconducting wires made out of GaAs or InAs are not free of disorder. A chain of adatoms on the surface also naturally introduces potential scattering terms. We introduce some quenched disorder $H_{\rm dis}=\int dx V(x) \rho(x)$
in $H_{el}$ where $\rho(x)$ is the electron density and $V(x)$ encodes  Gaussian disorder characterized by $\langle V(x)V(y)\rangle=D\delta(x-y)$ with $D$ the disorder strength.
Technically, disorder introduces extra  terms proportional to  $\cos\sqrt{2}(\phi_c\pm\phi_s)$ in the bosonized Hamiltonian \cite{giamarchi}. After disorder averaging, disorder terms generally competes with $B_{\rm eff}$ and $\Delta$.
As before, we assume again $B_{\rm eff}$ to be larger than $\Delta$. By comparing the scaling dimensions of the disorder and spiral magnetic field terms \cite{strom}, one finds that
when $(D a/4\pi v^2)> (2\Delta a/v)^\gamma$ with $\gamma=(3-K_c-K_s)/(2- (K_c+K_s^{-1})/2)$ (we assumed $v=v_c=v_s$ here), disorder dominates and ultimately leads to localization. However, when  $B_{\rm eff}$ is the largest energy scale, we can project the disorder term into the `$-$' helical state which renders the disorder term, being non-magnetic,  inoperant at lowest order in $D$ \cite{qi_rmp,note1}.
When all scales are of the same order, this is a difficult problem which goes beyond the scope of the present analysis. Therefore, the disorder energy scale much be at least smaller than the effective Zeeman field to observe the effect.

{\em Application.}
Let us first consider semiconducting GaAs or InAs wires. We take $L\sim 15\mu$m, $v_F=2\times 10^5{\rm ms}^{-1}$  and $k_F=10^8{\rm m}^{-1}$, and a proximity induced gap $\Delta_S\sim 0.2$meV \cite{mourik}. The hyperfine coupling for GaAs is $A_0=90\mu$eV and  $I=3/2$. For $K_c=1$, we obtain $B_{\rm eff}\sim 0.2$meV and  $T^*\lesssim 1$mK, too small to be observed.
More interesting is InAs which has $I=9/2$ and $A_0\sim 110\mu$eV \cite{coish}. For $K_c=0.8$, we obtain $T^*\sim 40$mK and $B_{\rm eff}^* \sim0.5{\rm meV}>\Delta^*\sim 0.1$meV (which guaranties a topological SC phase)  while for $K_c=0.6$, $\Delta^*\sim 200$mK and $T^*\sim 70$mK  which is within experimental reach.
For Co atoms on a Nb surface \cite{yazdani}, we take $I=5/2$, $a_I\sim 3$\AA, $E_F=v_Fk_F/2\sim 10$meV and $\Delta_S\sim 1$meV. For $K_c=1$, assuming a topological phase, {\it i.e.}, $B_{\rm eff}\sim IA_0 > \Delta_S$, we find that $T^*> \Delta_S/k_B$. Therefore a local magnetic coupling $A_0$ on the order of $0.5$meV, which is actually in the right range for magnetic exchange interactions \cite{menzel}, pushes the system in the topological phase.

{\em Conclusion.}
We have shown that in 1D a strong entanglement between magnetic moments and the electrons leads naturally
at low temperature to a magnetic spiral structure. Combined with a proximity induced superconducting gap,
this structure can drive the system into a topological superconducting phase supporting Majorana fermions.
A fine-tuning of external parameters is not required.
This scenario applies to semiconducting wires with nuclear spins or to a chain of magnetic adatoms on
top of a superconductor surface.
We also demonstrated that moderate electron interactions help stabilizing the topological phase.


{\em Acknowledgments.} We thank M. Franz, D. Loss and P. Wahl for helpful discussions and correspondences.
PS would like to thank the department of physics in UBC (Vancouver) for its kind hospitality during the final stage of this work. BB acknowledges the support by the EU-FP7 project SE2ND [271554].

{\em Note added.} During the final completion of this manuscript, we became aware of Ref. \cite{klinovaja,franz} which has
overlaps with the present work in the non-interacting limit.



\newpage


\section{Supplementary material}

\subsection{Susceptibility in the non-interacting helical superconductor}

In the non-interacting limit, the topological superconductor is described by an
effective $p$-wave BCS theory with the single-particle Hamiltonian
\begin{equation}
	H = {\sum_{k}}' \bigl(c_{k\up}^\dagger, c_{-k\dw}\bigr)
	\begin{pmatrix} \xi_k & \Delta \\ \Delta & -\xi_{-k} \end{pmatrix}
	\begin{pmatrix} c_{k\up} \\ c_{-k\dw}^\dagger \end{pmatrix},
\end{equation}
with the single electron energies $\xi_k = \xi_{-k}$ measured from the Fermi surface
and $c_{k\sigma}^\dagger$ the electron operators for momentum $k$ and spin $\sigma$.
We assume that the helical conductor is characterized by right moving modes with $\sigma=\up$
and left moving modes with $\sigma=\dw$. The $\Sigma'$ symbol marks a restriction of the momentum sum
to positive values $k \sim + k_F$.

A straightforward diagonalization of this model leads to the standard expressions
\begin{equation}
	H = E_0 + {\sum_k}' E_k (\alpha_k^\dagger \alpha_k + \beta_k^\dagger \beta_k),
\end{equation}
with $E_0$ the ground state energy, $E_k = \sqrt{\xi_k^2 + \Delta^2}$,
and the quasiparticle operators $\alpha_k, \beta_k$
related to the electron operators by
\begin{align}
	c_{k\up} &= u_{k} \alpha_k + v_{k} \beta_k^\dagger,
	\\
	c_{-k\dw}^\dagger &= v_{k} \alpha_k - u_{k} \beta_k^\dagger,
\end{align}
for the wave functions
\begin{equation}
	u_{k} = \sqrt{1 + \xi_k/E_k} / \sqrt{2}, \ \
	v_{k} = \sqrt{1 - \xi_k/E_k} / \sqrt{2}.
\end{equation}
The field operators for the left and right moving electrons with opposite spins
are then expressed as
\begin{align}
	\psi_\up(x) &= \frac{1}{\sqrt{N}} {\sum_{k}}' \e^{ikx} \left[ u_{k} \alpha_k + v_{k} \beta_k^\dagger\right],
\\
	\psi_\dw(x) &= \frac{1}{\sqrt{N}} {\sum_{k}}' \e^{-ikx} \left[ v_{k} \alpha_k^\dagger - u_{k} \beta_k \right],
\end{align}
for $N$ the number of sites.
We have defined the
transverse electron spin susceptibility as
\begin{align}
	\chi^{\perp}(q)
	&= -\frac{i}{a_I^2} \int_0^\infty dt \int dx \ \e^{i q x} \ \mean{[S_+(x,t),S_-(0,0)]}
\nonumber\\
	&= \frac{1}{Na_I} {\sum_{k}}'
	\frac{2 u_{k} u_{q-k} v_{k} u_{q-k} - u_{k}^2 v_{q-k}^2 - v_{k}^2 u_{q-k}^2}{E_k+E_{q-k}}.
\label{eq:chi(q)_interm}
\end{align}
In the latter sum, $k$ as well as $q-k$ are restricted to positive values close to $k_F$, which we
can capture by introducing a momentum cutoff such that $k, q-k > \Lambda$ with $\Lambda \ll k_F$.
The summation in Eq. \eqref{eq:chi(q)_interm} is therefore restricted to $\Lambda < k < q-\Lambda$.
The latter sum leads to a well marked minimum of $\chi^\perp(q)$ at $q=2k_F$.
In the regime $\xi_k \approx v_F (k-k_F)$ and for $\Delta/ v_F k_F \ll 1$,
we can let $\Lambda \to 0$. We furthermore go over to continuous $k$, which allows us 
to straightforwardly integrate Eq. \eqref{eq:chi(q)_interm}, and we obtain
with $\bar{q} = q - 2k_F$ and $\eta_k = v_F k$
\begin{align}
	&\chi^{\perp}(q)
	=
	\frac{-1}{8\pi v_F}
	\Biggl\{
		\ln\Biggl[
			\frac{\eta_{k_F}+\sqrt{\Delta^2+\eta_{k_F}^2}}{-\eta_{k_F}+\sqrt{\Delta^2+\eta_{k_F}^2}}
		\Biggr]
\nonumber\\
		&+
		\ln\Biggl[
			\frac{\eta_{k_F}+\eta_{\bar{q}}+\sqrt{\Delta^2+(\eta_{k_F}+\eta_{\bar{q}})^2}}
			     {-\eta_{k_F}-\eta_{\bar{q}}+\sqrt{\Delta^2+(\eta_{k_F}+\eta_{\bar{q}})^2}}
		\Biggr]
		+
		2
		\frac{\sqrt{4 \Delta^2+\eta_{\bar{q}}^2}}{\eta_{\bar{q}}}
\nonumber\\
	&\times
		\ln\Biggl[
			\frac{2 \Delta^2 - \eta_{\bar{q}}\eta_{k_F}         +\sqrt{\Delta^2+\eta_{k_F}^2        } \sqrt{4 \Delta^2+\eta_{\bar{q}}^2}}
			      {2 \Delta^2 + \eta_{\bar{q}}\eta_{k_F+\bar{q}}+\sqrt{\Delta^2+\eta_{k_F+\bar{q}}^2} \sqrt{4 \Delta^2+\eta_{\bar{q}}^2}}
		\Biggr]
	\Biggr\}.
\end{align}
This function has a minimum at $q=2k_F$ given by
\begin{align}
	&\chi^\perp(2k_F)
	= 
\\
	&\frac{-1}{4 \pi v_F} 
	\Biggl\{
		\ln\Biggl[
			\frac{ \eta_{k_F} +\sqrt{\Delta^2+\eta_{k_F}^2}}
			      {-\eta_{k_F} +\sqrt{\Delta^2+\eta_{k_F}^2}}
		\Biggr]
		+
		\frac{2 \eta_{k_F}}{\Delta+\sqrt{\Delta^2 + \eta_{k_F}^2}}
	\Biggr\}.
\end{align}
In the regime $\Delta/\eta_{k_F} \ll 1$ and for $\bar{q}\ll k_F$ such that 
$\eta_{k_F + \bar{q}} \approx \eta_{k_F}$ we can expand these results to the 
more speaking expressions
\begin{align}
	&\chi^\perp(q)
	\approx
	\frac{-1}{2\pi v_F}
	\Biggl[
		\ln\left(\frac{2 v_F k_F}{\Delta}\right)
		-
		\frac{1}{\kappa}
		\arctanh(\kappa)
	\Biggr],
\end{align}
where $\kappa = \eta_{\bar{q}} / \sqrt{4 \Delta^2 + \eta_{\bar{q}}^2}$,
and 
\begin{equation}
	\chi^\perp(2k_F)
	\approx 
	\frac{-1}{2\pi v_F} \left\{ \ln\left[\frac{2 \eta_{k_F}}{\Delta} \right] - 1 \right\}.
\end{equation}
In the limit $|\eta_{\bar{q}}| \ll \Delta$ a further expansion of the $\arctanh$
leads to a Lorentzian shaped susceptibility
\begin{align}
	&\chi^\perp(q)
	\approx
	\bar{\chi}
	- \frac{1}{3\pi v_F}
	\frac{2 \Delta^2}{4 \Delta^2 + \eta_{\bar{q}}^2},
\end{align}
with $\bar{\chi} = - [ \ln\left(2 \eta_{k_F}/\Delta\right) - 4/3]/2\pi v_F$.

\subsection{Bosonized form of Hamiltonians}

The bosonization of the Hamiltonians follows the standard approach \cite{giamarchi}. 
The interacting electron Hamiltonian (in the absence of $B$ field and pairing amplitude)
is rewritten in terms of the boson fields $\phi_\nu,\theta_\nu$ for 
charge ($\nu=c$) and spin ($\nu=s$) degrees of freedom as
\begin{equation}
	H^0_{\nu} = \int \frac{dx}{2\pi} \left[
		\frac{v_\nu}{K_\nu} (\nabla \phi_\nu)^2 + v_\nu K_\nu (\nabla \theta_\nu)^2
	\right],
\end{equation}
for velocities $v_\nu$ and Luttinger liquid parameters $K_\nu$ as described in the main text.
In terms of the boson fields, the electron operators $\psi_{r\sigma}(x)$ for 
right ($r=+$) and left ($r=-$) moving modes and spin $\sigma = \up,\dw= +,-$ is given by 
$\psi_{r\sigma} = (\eta_{r\sigma}/\sqrt{2\pi a}) \exp(i r k_F x) 
\exp(-i [r \phi_c - \theta_c + \sigma (r \phi_s - \theta_s)] /\sqrt{2})$,
where $a$ is the short distance cutoff, and $\eta_{r\sigma}$ the Klein factors taking 
into account the fermion statistics and satisfying 
$\{\eta_{r\sigma}^\dagger,\eta_{r'\sigma'}\} = \delta_{r,r'} \delta_{\sigma,\sigma'}$. 

For a spiral magnetic field $\bB(x) = B (\cos(2k_F x), \sin(2k_F x), 0)$, we use the
bosonized form of the electron operators to rewrite the Zeeman coupling as
\cite{braunecker:2009b}
\begin{align}
	\bB(x) \cdot \bS(x)
	= \frac{B}{4\pi a} 
	\Bigl[ 
		&\cos(\sqrt{2}(\phi_c + \theta_s)) 
\nonumber\\
		+ 
		&\cos(\sqrt{2}(\phi_c - \theta_s) - 4 k_F x) 
	\Bigr].
\label{eq:HZ_boson}
\end{align}
We have dropped eventually the Klein factors as they do not have any further influence.
As long as $k_F$ is not commensurate with the lattice the second term in Eq. \eqref{eq:HZ_boson} 
oscillates quickly in space
and its contribution to the Hamiltonian is irrelevant. The first term, however, is relevant and 
provides Eq. (6) in the main text.

In the same way, the pairing amplitude is bosonized for terms 
$\Delta \psi_{r\sigma}^\dagger \psi_{-r,-\sigma}^\dagger + h.c.$ and leads to Eq. (7) 
in the main text.

\subsection{Change of basis}

We introduce the new boson fields
$\phi_+$, $\phi_-$, $\theta_+$ and $\theta_-$, are defined  as follows \cite{braunecker:2009b},
\begin{eqnarray}
\phi_c &=& \sqrt{\frac{K_c}{K}}\Big[  \sqrt{K_c} \phi_+   -\frac{1}{\sqrt{K_s}}  \phi_-\Big ] {}
\nonumber\\
\phi_s &=& -\sqrt{\frac{K_s}{K}}\Big[  \frac{1}{\sqrt{K_s}} \theta_+   + \sqrt{K_c}  \theta_-\Big ] {}
\nonumber\\
\theta_c &=& -\frac{1}{\sqrt{K_c K}}\Big[  \sqrt{K_c} \theta_+   -\frac{1}{\sqrt{K_s}}  \theta_-\Big ] {}
\nonumber\\
\theta_s &=& \frac{1}{\sqrt{K_s K}}\Big[  \frac{1}{\sqrt{K_s}} \phi_+   + \sqrt{K_c}  \phi_-\Big ], \label{eq:newbasis}
\end{eqnarray}
where $K=K_c + K_s^{-1}$ and the fields obey the standard commutation relations $[\phi_a(x),\theta_b(y')]=i\pi\delta_{ab}\text{sign}(x-y)$ with $a,b=\pm$.



\begin{thebibliography}{99}
\bibitem{note1} This becomes here manisfest as terms
$\propto D\cos(\kappa_1\phi_-)\cos(\kappa_2\theta_-)$
which are of zero average, where $\kappa_{1,2}$ are unimportant functions of $K_c,K_s$.
\end{thebibliography}
\end{document}